# Valley selective optical control of excitons in 2D semiconductors using Chiral metasurface

S. Guddala[1], R. Bushati[1,2], M. Li[2,3], A. B. Khanikaev[2,3], and V. M. Menon[1,2,*]

[1] Department of Physics, City College, City University of New York (CUNY), New York, USA

[2] Department of Physics, Graduate Center, City University of New York (CUNY), New York, USA

[3] Department of Electrical Engineering, City College, City University of New York (CUNY), New York, USA

*vmenon@ccny.cuny.edu

**Abstract:** Recent advances in condensed matter physics have shown that the valley degree of freedom of electrons in 2D materials with hexagonal symmetry, such as graphene, h-BN, and TMDs, can be efficiently exploited, leading to the emergent field of valleytronics, which offers unique opportunities for efficient data transfer, computing and storage. The ability to couple the valley degree of freedom of electrons with light can further expand the ways one manipulate this degree of freedom, thus envisioning a new class of solid-state-photonic interfaces and devices. Besides this expansion of control of valley by light-waves, coupling of photons with valley-polarized electrons can dramatically expand the landscape of available optical responses, which may bring new means of controlling light in photonic devices. In this work we design such hybrid solid-state photonic metasurface integrating 2D TMD and photonic all-dielectric metasurface. While TMD is naturally endowed with the property of valley to optical-polarization coupling, the photonic metasurface is designed to produce chiral field which selectively couples to the valley degree of freedom of solid-state TMD component. We experimentally demonstrate that such coupling leads to controlled valley polarization due to the coupling of 2D materials with the chiral photonic metasurface. The measured emission from valley excitons in this hybrid system yields the preferential emission of specific helicity.

## 1. Introduction

Atomic monolayers of transition metal dichalcogenides (TMDs) have seen large scientific interest for the past few years, owing to their dramatic quantum confinement effects, such as indirect to direct band gap transition, broken inversion symmetry, large exciton binding energy and strong photoluminescence. In addition, the broken inversion symmetry in monolayer TMDs result in valley specific light emission due to the excitation of a specific valley in the momentum space, i.e, the optical transitions at the K and K' valleys are allowed for $\sigma^+$ (right) and $\sigma^-$ (left) circular polarized lights, respectively – the phenomenon referred to as "valley polarization". The generation and detection of valley polarization has been achieved by pumping directly light with specific handedness [1–3]. While the K and K' valleys have different selection rules, they are energetically degenerate and this degeneracy was lifted using the valley Zeeman effect using either magnetic fields [4,5] or through intense optical pumping[6]. However, all of these demonstrations were far from an integrated chip-scale architecture. The control and manipulation of valley specific transitions (K or K') has been in a strong demand for its application in spintronics, valleytronics and as information carriers in quantum computation. More recently significant interest emerged in the context of photonics, and coupling of plasmons, which naturally have chiral properties [7,8], to excitons in TMDs was investigated [9–11].

Here, we report a novel approach to address selective valley polarization at room temperature by engineering a hybrid solid-state photonic interface, whose photonic component represents a dielectric Chiral metasurface generating chiral near field for the linear polarized incidence. This Chiral metasurface is interfaced with $WS_2$ TMD monolayer and generates left and right handed circular polarization depending on the orientation of the incident linear polarization. Thus, the metasurface design permits selective valley (K and K') excitation, and respective $\sigma^+$ (left) and $\sigma^-$ (right) circularly polarized light emission from the $WS_2$ TMD monolayer and the hybrid device as a whole. So, the designed metasurface has shown the ability to generate preferred handedness of the circular polarization due to the excitation of a specific valley in the $WS_2$ monolayer.

## 2. Chiral metasurface design and simulations

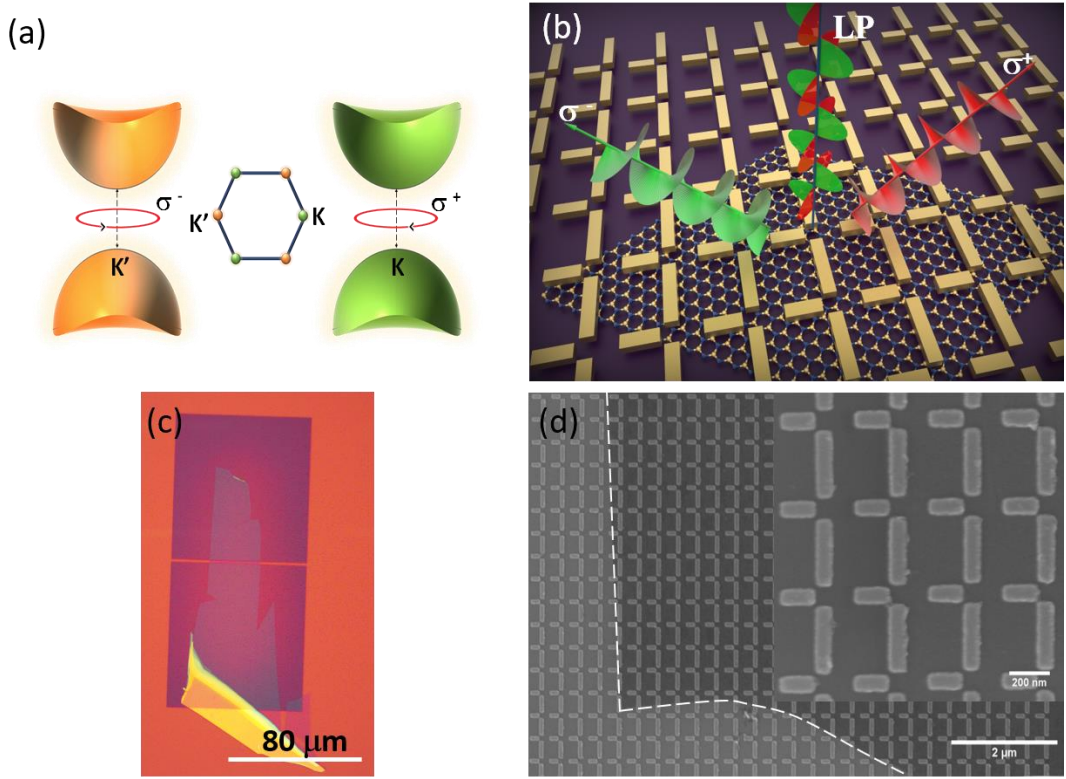

Figure 1. a) Schematic of the valley polarization in TMD monolayer. b) Schematic of the Chiral metasurface showing the linear polarization control on valley selective polarization. c) Optical image of the structure with WS$_2$ monolayer. d) SEM image of Ge Chiral metasurface, inset shows high magnification image of the Chiral metasurface. The white dashed line indicates the boundary of the WS$_2$ monolayer below the Ge metasurface.

The dielectric chiral metasurface is designed with two Ge nanorods of asymmetric lengths arranged orthogonal to each other on a WS$_2$ monolayer, which was mechanically exfoliated on to PDMS stamp and transferred to thermal oxide coated silicon substrate. Fig. 1b shows the schematic of achieving selective valley emission from the WS$_2$ monolayer for the linear polarization incidence on the chiral metasurface. The phase difference between reflected waves from the two orthogonal nanorods is tailored to be $90^0$ at the desired wavelength of operation by controlling the length, width and thickness of the nanorods as well as the gap between them [12]. At the wavelength between the resonances of individual antennas the metasurface generates right handed circular chiral nearfield for the linearly polarized incidence of $+45^0$ with respect to the long rods and similarly generates left handed circular chiral near field for $-45^0$, as schematically shown in Fig.1b.

This nearfield configuration stems from the linear superposition of two dipoles interfering in the hotspots near the tips of the antennas, which, due to $90^0$ phase shift, result in circularly polarized field profile of definite handedness (Fig. 2). The handedness of the near-field chirality can be controlled by the polarization of incident light, and it can be reversed by changing the incident polarization to $-45^0$ (Fig.2).

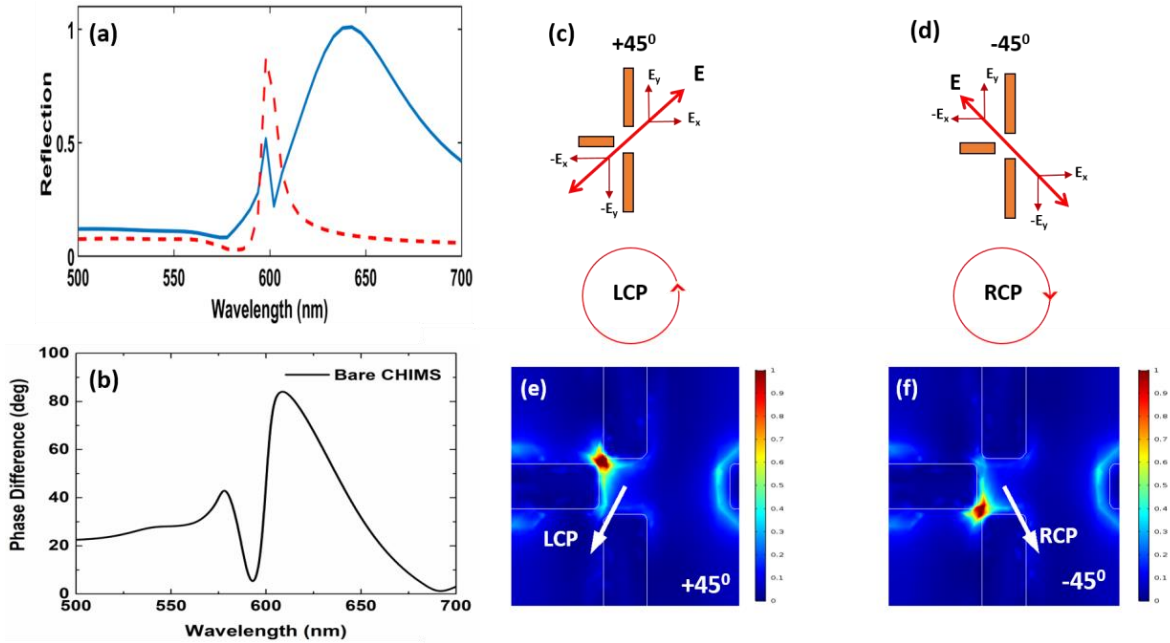

**Figure 2: (a)** Dipole resonances of longer (blue) and shorter rod (red) for the linear polarization incidence along their longer axes. b) Near $90^0$ phase difference resulting from linearly polarized light incident at $+45^0$ with respect to both orthogonal rods as shown in (c). Schematic unit cell of the Chiral metasurface illustrating the linear polarization electric field excitation at (c) $+45^0$ and (d) $-45^0$ with respect to the orthogonal rods, and the corresponding circular polarization with preferential handedness that is generated. Simulated electric field distributions arising from the two antenna for linear polarized excitation resulting in left (e) and right (f) circularly polarized light, respectively. Media files attached to the simulations show circular polarization of definite handedness from the metasurface for linear polarization incidence.

The metasurface was designed and the structural dimensions were optimized with the use of the finite element method (FEM) based software COMSOL Multiphysics. The resultant unit cell of the structure contains two Ge rods that are 165 nm and 285 nm, which are aligned in horizontal and vertical directions, respectively, and have a corner spacing of 20 nm. The width and height of the rods are 65 and 45 nm respectively. The simulated dipolar resonant wavelengths for short rod and long rod are at 598 nm and 648 nm, respectively. A linearly polarized light incidence at $+45^0$ with respect to the two orthogonal rods can simultaneously excite the rods to resonate at their resonant frequencies. But given the $90^0$ phase shift between the resonances of the two rod, circular polarization of definite handedness is produced (Fig. 2c). Similarly, the $-45^0$ incidence results in opposite handed circular polarization with respect to $+45^0$ incidence (Fig. 2d). The handedness of the circular polarization can be summarized from the phase difference between the $E_x$ and $E_y$ components of linear polarization for $+45^0$ and $-45^0$ incidences as shown in Fig. 2b. This is an indication of circular polarization generation from this orthogonal geometry of two asymmetric nanorods at the choice of wavelength. The optimized nanorod geometry shows this chiral resonance at $WS_2$ exciton resonance wavelength of 615 nm. The electric field distribution (Fig. 2e and Fig. 2f) show hotspots with specific handed circular polarization. The electric field distribution corresponding to the resonance at 625 nm shows strong confinement of electric field at the 20 nm gap between the rods and between the antennas, revealing chiral hotspots of left and right handedness. The media files attached to the Fig. 2e and Fig. 2f show the circular polarization response from the Chiral metasurface for the linear polarization incidence.

## 3. Experimental results

### 3.1. Fabrication:

An atomic-layer-thick monolayer of $WS_2$ was mechanically exfoliated onto a PDMS stamp and transferred to the 300 nm thick thermal oxide deposited silicon substrate. The monolayer was annealed in the nitrogen atmosphere at $300^0C$ prior to polymer spin coating. The dielectric chiral metasurface (Fig. 1d) was fabricated with the use of e-beam lithography. A positive photoresist 495-PMMA-A4 (Microchem) was spin coated at 3000 rpm and annealed at $180^0C$ for 2 min. The chiral metasurface pattern was written with finely focused electron beam followed by the development in isopropanol and water mixture. A 40 nm thick Ge was deposited by e-beam assisted physical evaporation system. The residual PMMA was removed by acetone lift-off process. An optical image of the sample can be seen in the Fig. 1c. Large area uniformity of the metasurface can be seen in the SEM image of the metasurface shown in Fig. 1d. The inset of the figures indicate the fine rod shapes of the structures with 20 nm corner spacing between orthogonally arranged long and short rods.

## 3.2. Optical characterization

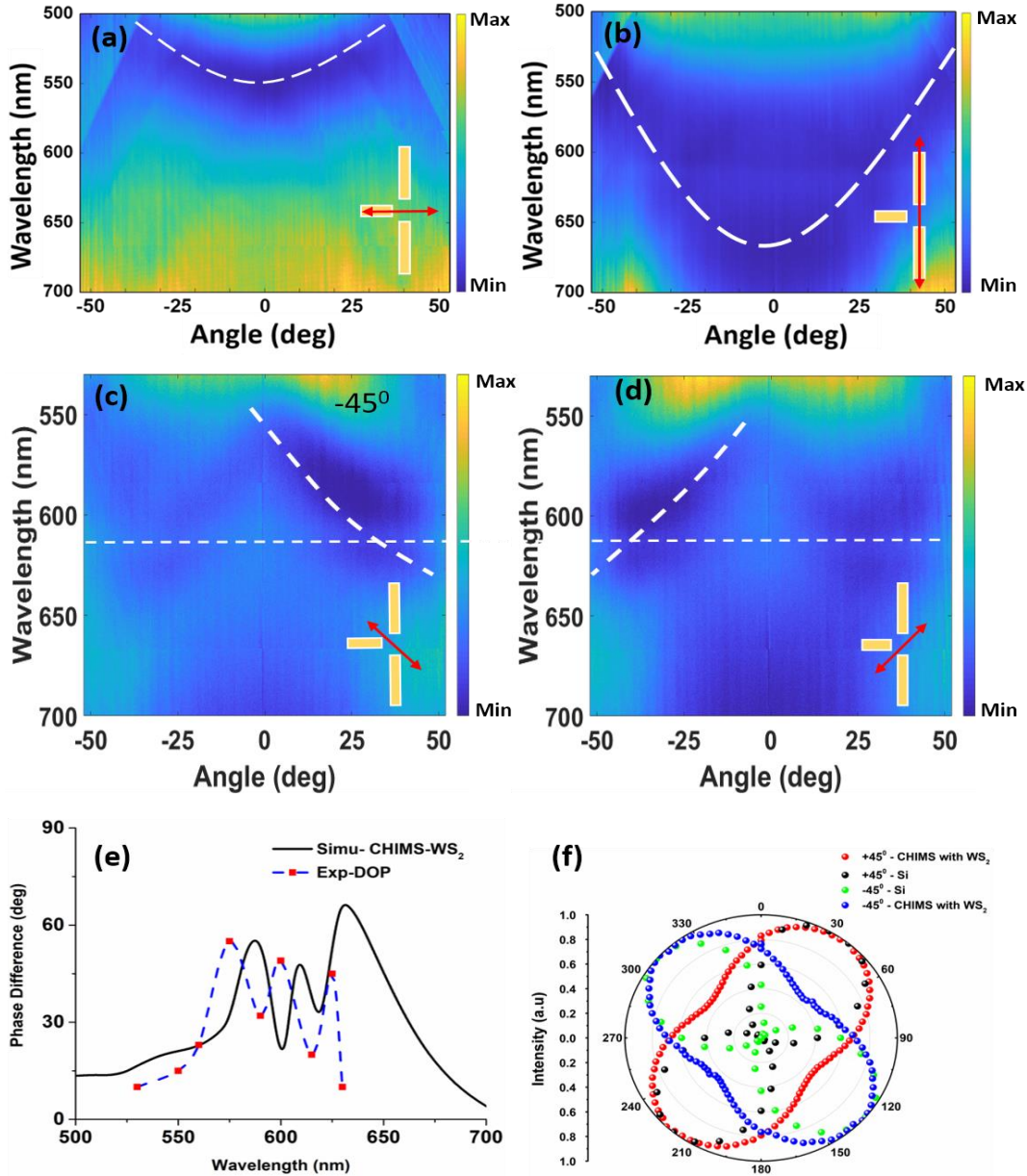

**Figure 3:** (a,b) Dipolar resonances for the linearly polarization incidence with in-plane polarization along (a) vertical and (b) horizontal rods. (c,d) Similar spectra for the linearly-polarized incidence with in-plane polarization at (c) $+45^0$ and (d) $-45^0$ with respect to rods. White dashed lines are guide to eye to show the dipolar and chiral modes of resonances. e) Wavelength dispersion of simulated (black line) degree of polarization for CHIRAL metasurface with $WS_2$ monolayer. Wavelength dispersion of degree of circular polarization measured (red circles) at different wavelengths of excitation. Blue dashed line is a guide to eye. f) Reflected laser intensity from the CHIRAL metasurface with $WS_2$ for the excitation wavelength of 575 nm as a function of angle of rotation of the second linear polarizer (analyzer) for $-45^0$ (blue) and $+45^0$ (red) incidence of in-plane linear polarization. Similarly, the black and green circles are measurement on reference bare silicon substrate.

The dispersion of the structure (Fig. 3) was measured using Fourier space (k-space) imaging spectroscopy with linearly polarized white light. The structure shows dipolar resonances for linear polarization along short and long Ge rods at 540 nm and 680 nm wavelengths with dispersions as shown in Fig.3a and 3b, respectively. For linear polarization at $-45^0$ angle as shown in Fig.3c, an asymmetric dispersion towards +k direction is observed. In contrast, for $+45^0$ angle of incidence this dispersion is found to be flipped towards -k direction as shown in Fig. 3d. The polarization state of the reflected light is measured using tunable laser producing a monochromatic linear polarized source. A second linear polarizer is placed between the sample and the detector and rotated from $0^0$ to $360^0$ in steps of $4^0$ to analyze the angular-dependence of the reflected light at each wavelength.

The degree of linear polarization (DOP) is measured from the maximum to minimum ratio of reflected laser intensities. The DOP, as shown in Fig. 3e, is found to be maximum at wavelengths 575 nm, 600 nm and 625 nm as 55%, 50% and 47% respectively. These experimental results agree with the simulation results of chiral metasurface with $WS_2$ monolayer shown by black colored curve in Fig. 3e. However, the experimentally measured resonances are found be differ with simulations by 10 nm, which is attributed to the imperfections in the nanorod dimensions due to e-beam scattering during lithography process. The reduced degree of circular polarization of chiral metasurface with $WS_2$ in comparison to the bare chiral metasurface is found to be due to the absorption losses of $WS_2$ monolayer. Fig. 3f shows the reflected laser intensity at 575nm wavelength of excitation as a function of the angle of the second polarizer. The elliptical shape of the angular dependence of the reflected light suggests that the output light is circularly polarized to a large degree in far field, thus confirming chiral near-field. The observations for $-45^0$ angle of incidence is found to be flipped as shown in Fig. 3f.

## 4. Results and Discussion

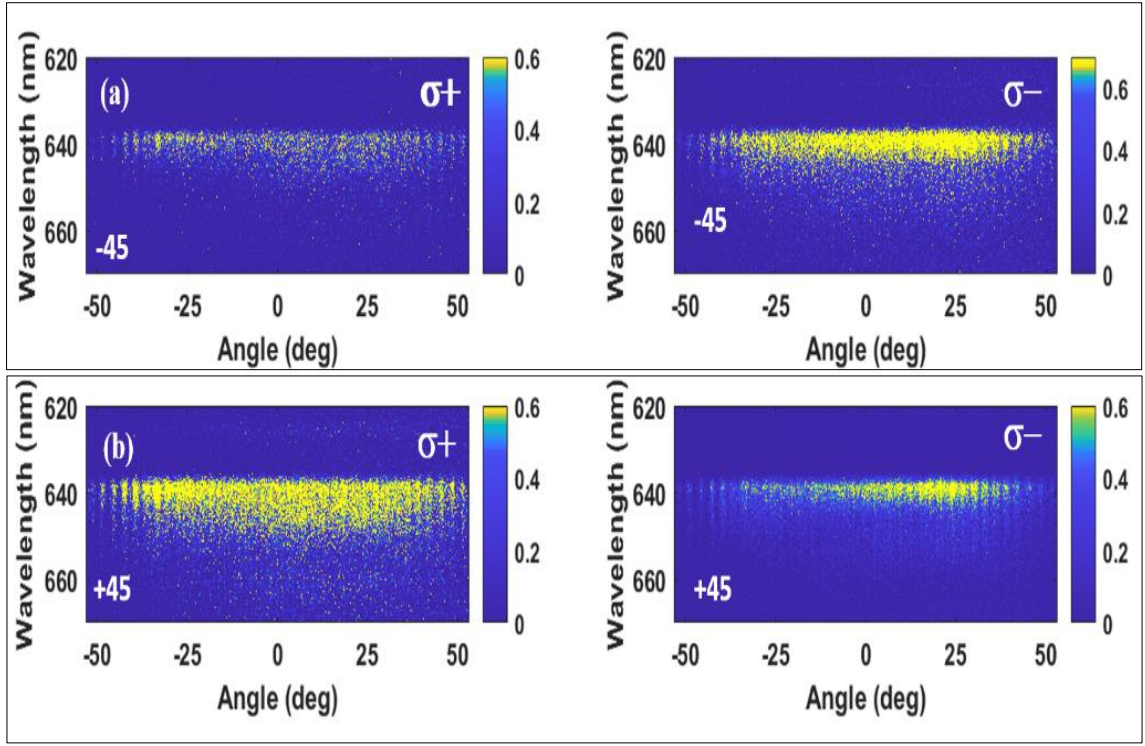

**Figure 4:** k-space PL dispersions for the wavelength of excitation of 625 nm are resolved into σ+ (left handed) and σ- (right handed) circular polarization emissions for the incidence of linear polarization at (a) -45$^0$ and (b) +45$^0$, as shown in the unit cell in Fig. 2.

The selective valley polarization of the $WS_2$ monolayer is investigated by fabricating the chiral metasurface on the top of the monolayer as shown in Fig. 1d. Having the monolayer on top of the metasurface is found to have a detrimental effect on valley polarization due to strain induced intervalley scattering [13]. Therefore, the chiral metasurface was fabricated on top of the monolayer. The experimental observations were performed by using tunable pulsed laser source (Toptica) and k-space spectroscopy. Valley polarization of the bare monolayer of $WS_2$ without the metasurface was studied with excitation wavelength of 625 nm for circular polarization incidence. The photoluminescence (PL) emission from the bare $WS_2$ monolayer on excitation with σ $^+$ (LCP) and σ $^-$ (RCP) polarizations were resolved respectively into σ $^+$ (left handed) and σ $^-$ (right handed) circular polarization basis by using the combination of a quarter wave plate and a linear polarizer. The laser line was filtered by using a long pass 635 nm filter. The emission profiles were dispersed by 500 gr/mm grating, Princeton Instruments monochromator, and analyzed by an air-cooled CCD detector. The metasurface with monolayer was excited with excitation wavelength of 625 nm with linear polarization at +45⁰ orientations to the two Ge rods as shown in Fig. 2. The structure is found to show 47% maximum degree of circular polarization at this wavelength. The k-space PL dispersion plots are found to show (Fig. 4a) minimum intensity for σ⁺ (left handed) and maximum for σ⁻ (right handed) circular polarization emissions at helicity resolved angles. These emission intensities are found to be reversed when the incident linear polarization is changed from +45⁰ to

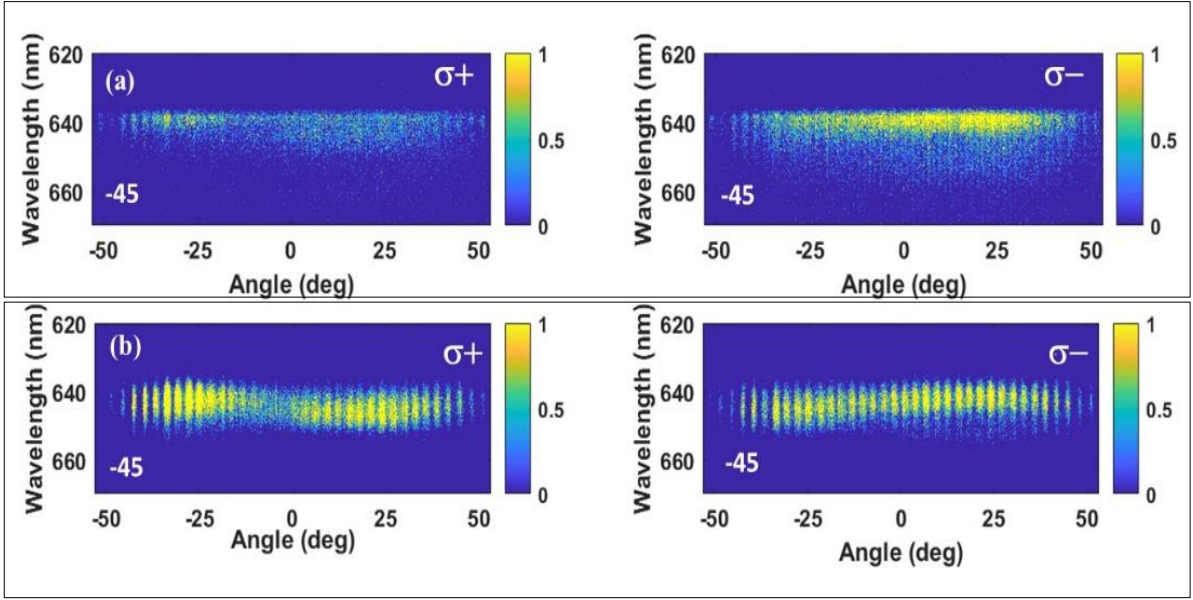

-45⁰ (Fig. 4b).

**Figure 5:** k-space PL dispersions resolved into σ+ (left handed) and σ- (right handed) circular polarization emissions for the incidence of linear polarization at -45⁰, as shown in the unit cell in Fig. 2 for the wavelengths of excitation (a) 600 nm, (b) 575 nm.

Similar measurements for the excitation at chiral metasurface structural resonance of 600 nm is found to be similar but with decreasing contrast between σ⁺ (left handed) and σ⁻ (right handed) circular polarization emissions for -45⁰ as shown in Fig. 5a. Similarly, for the wavelength of excitation at 575 nm, far from the $WS_2$ exciton resonance shows almost equal intensities for both σ⁺ (left handed) and σ⁻ (right handed) circular polarization emissions. The observations for +45⁰ are found to be reversed for the two cases similar to the 625 nm excitation (not shown here). These observations indicate that the inter-valley scattering is predominant at room temperature for the excitations above the exciton resonance, in agreement with the earlier studies on bare TMD materials[14]. In contrast, the excitation at 625 nm which is very close to the exciton resonance, shows less inter-valley scattering and favors achieving linear polarization control on valley selective optical emission.

K and K' valleys emission contrast can be defined by measuring the optical helicity given by

$$\rho = \frac{I(\sigma^-) - I(\sigma^+)}{I(\sigma^-) + I(\sigma^+)}$$

where $I(\sigma^-)$ and $I(\sigma^+)$ are intensities corresponding to the $\sigma^-$ and $\sigma^+$ emissions for linear polarization excitation on chiral metasurface. At 625 nm wavelength linear polarization excitation, the emission from $WS_2$ shows optical helicity ($\rho$) of 27% although the linear to circular conversion of the excitation is not as efficient at this wavelength. However, the contrast between the emissions for the two polarizations is large because of the suppressed intervalley scattering. Though the structure at 600 nm excitation preserves high degree of circular polarization, this excitation is above the exciton resonance and hence the intervalley scattering dominates resulting in lower contrast valley polarization with optical helicity ($\rho$) of 20%. Similarly the intervalley scattering gets even more dominant for the 575 nm excitation which is far above the exciton resonance.

In summary, we have shown the possibility to selectively address the specific valleys in 2D TMDs using a Chiral metasurface that generates hotspots with specific handedness. Despite the highly spatially localized nature of these resonances, the overall response of the metasurface shows enhanced degree of circular polarization. The use of such Chiral metasurface provides a platform for developing chip-scale valleytronic devices wherein the valley degree of freedom can be accessed optically.

## 5. Acknowledgements

Authors acknowledge financial support of Army Research Office through grant number W911NF-16-1-0256 and NSF through grant number DMR-1709996 and DARPA Nascent Light-Matter Interaction program. Author S. Guddala acknowledge the fellowship from IUSSTF and SERB India. This work was performed in part at the Advanced Science Research Center (ASRC) NanoFabrication Facility of the Graduate Center at the City University of New York.

**References**
1.    Mak, K. F., He, K., Shan, J. & Heinz, T. F. Control of valley polarization in monolayer

MoS2by optical helicity. *Nat. Nanotechnol.* **7,** 494–498 (2012).

2. Mak, K. F., McGill, K. L., Park, J. & McEuen, P. L. The valley Hall effect in MoS2 transistors. *Science (80-. ).* **344,** 1489–1492 (2014).
3. Cao, T. *et al.* Valley-selective circular dichroism of monolayer molybdenum disulphide. *Nat. Commun.* **3,** 885–887 (2012).
4. Srivastava, A. *et al.* Valley Zeeman effect in elementary optical excitations of monolayer WSe2. *Nat. Phys.* **11,** 141–147 (2015).
5. Macneill, D. *et al.* Breaking of valley degeneracy by magnetic field in monolayer MoSe2. *Phys. Rev. Lett.* **114,** 1–5 (2015).
6. Sie, E. J. *et al.* Valley-selective optical Stark effect in monolayer WS2. *Nat. Mater.* **14,** 290–294 (2015).
7. Bliokh, K. Y., Smirnova, D. & Nori, F. Quantum spin Hall effect of light. *Science (80-. ).* **348,** 1448–1451 (2015).
8. Van Mechelen, T. & Jacob, Z. Universal spin-momentum locking of evanescent waves: supplemental document. *Optica* **3,** 1–10 (2016).
9. Gong, S. H., Alpeggiani, F., Sciacca, B., Garnett, E. C. & Kuipers, L. Nanoscale chiral valley-photon interface through optical spin-orbit coupling. *Science (80-. ).* **359,** 443–447 (2018).
10. Chervy, T. *et al.* Room Temperature Chiral Coupling of Valley Excitons with Spin-Momentum Locked Surface Plasmons. *ACS Photonics* **5,** 1281–1287 (2018).
11. Sun, L. *et al.* Routing Valley Excitons in a Monolayer MoS2 with a Metasurface. *arXiv:1801.06543* (2018).
12. Zhao, Y. & Alù, A. Manipulating light polarization with ultrathin plasmonic metasurfaces. *Phys. Rev. B - Condens. Matter Mater. Phys.* **84,** 1–6 (2011).
13. Zhu, C. R. *et al.* Strain tuning of optical emission energy and polarization in monolayer and bilayer MoS2. *Phys. Rev. B - Condens. Matter Mater. Phys.* **88,** 1–5 (2013).
14. Nayak, P. K., Lin, F. C., Yeh, C. H., Huang, J. S. & Chiu, P. W. Robust room temperature valley polarization in monolayer and bilayer WS2. *Nanoscale* **8,** 6035–6042 (2016).